\documentclass[11pt]{article}

\usepackage{amsmath, amssymb, amsthm, bm, mathtools}
\usepackage{geometry}
\usepackage{setspace}
\usepackage{natbib}
\usepackage{graphicx}
\usepackage{booktabs}
\usepackage{array}
\usepackage{enumitem}
\usepackage{algorithm}
\usepackage{algpseudocode}
\usepackage{hyperref}
\usepackage{xcolor}

\hypersetup{
    colorlinks=true,
    linkcolor=blue!60!black,
    citecolor=blue!60!black,
    urlcolor=blue!60!black
}

\newcommand{\R}{\mathbb{R}}
\newcommand{\Sph}{\mathbb{S}}

\newcommand{\Corr}{\operatorname{Corr}}
\newcommand{\diag}{\operatorname{diag}}
\newcommand{\logit}{\operatorname{logit}}

\newcommand{\offdiag}{\operatorname{offdiag}}

\newcommand{\bfy}{\mathbf{y}}
\newcommand{\bfx}{\mathbf{x}}

\newcommand{\bfw}{\mathbf{w}}
\newcommand{\bfh}{\mathbf{h}}
\newcommand{\bff}{\mathbf{f}}
\newcommand{\bfu}{\mathbf{u}}

\newcommand{\bfs}{\mathbf{s}}
\newcommand{\bfB}{\mathbf{B}}
\newcommand{\bfC}{\mathbf{C}}
\newcommand{\bfD}{\mathbf{D}}

\newcommand{\bfI}{\mathbf{I}}

\newcommand{\bfQ}{\mathbf{Q}}
\newcommand{\bfR}{\mathbf{R}}
\newcommand{\bfU}{\mathbf{U}}
\newcommand{\bfV}{\mathbf{V}}

\newcommand{\bfY}{\mathbf{Y}}

\newcommand{\bfzero}{\boldsymbol{0}}
\newcommand{\bfone}{\bm{1}}
\newcommand{\bfalpha}{\bm{\alpha}}
\newcommand{\bfbeta}{\bm{\beta}}
\newcommand{\bfgamma}{\bm{\gamma}}
\newcommand{\bfdelta}{\bm{\delta}}
\newcommand{\bfeta}{\bm{\eta}}
\newcommand{\bfepsilon}{\bm{\epsilon}}
\newcommand{\bflambda}{\bm{\lambda}}
\newcommand{\bfmu}{\bm{\mu}}
\newcommand{\bfomega}{\bm{\omega}}
\newcommand{\bfLambda}{\bm{\Lambda}}
\newcommand{\bfDelta}{\bm{\Delta}}

\theoremstyle{remark}

\title{Composition as Direction: An Active-Set Ray-Based Model for Sparse High-Dimensional Compositional Data}
\author{
  Michael R. Schwob$^{1}$\thanks{Corresponding author: \texttt{schwob@vt.edu}},\quad Jyotishka Datta$^{1}$ \\[4pt]
  {\small $^{1}$\textit{Department of Statistics, Virginia Tech}}
}
\date{\today}

\begin{document}

\maketitle

\begin{abstract}
Compositional data are central to microbial, ecological, and environmental research, yet often have four features that are difficult to accommodate jointly: exact zeros, latent dependence among components, high-dimensionality, and a unit-sum constraint that induces a non-Euclidean geometry. Conventional Dirichlet-type and logistic-normal models address these features only partially; the former imposes an inflexible covariance structure, and the latter requires strictly positive observations or auxiliary zero-handling approaches that are problematic under sparsity. Projected Gaussian models offer a directional representation that captures exact zeros and latent dependence; however, support correctness on the simplex requires either truncation or folding, both of which become computationally prohibitive as the dimension grows. We develop an Active-set Ray-based Compositional (ARC) framework, which retains the benefits of projected Gaussian models while remaining computationally feasible in high-dimensional settings. In this framework, we map compositions to the nonnegative orthant of the unit hypersphere and specify an active-set process that governs which components are present. Conditional on the active set, the positive subcomposition is modeled by evaluating a latent Gaussian density along positive rays of the active subspace with the radius treated as an auxiliary variable. Such a construction (i) separates the active-set process that governs which components are present from the positive subcomposition on the active components, (ii) preserves a latent Gaussian interpretation, and (iii) accommodates arbitrary latent dependence. 
Thus, the framework is conducive to high-dimensional applications in which exact zeros and shared positive responses are scientifically central.
Conceptually, the proposed framework reframes a composition as an observed direction of a latent abundance vector with an unobserved magnitude and an explicitly modeled active set.
\end{abstract}

\section{Introduction}
\label{sec:intro}

Compositional data arise across the environmental and life sciences: microbial communities \citep{gloor2017microbiome, quinn2018understanding}, phytoplankton assemblages \citep{pawlowsky2015tools}, soil and sediment microbiomes \citep{fierer2017embracing}, dietary intake records \citep{leite2017compositional}, and remotely sensed land-cover proportions \citep{leininger_spatial_2013} are all expressed naturally as compositions. Formally, a $d$-dimensional composition $\mathbf{u}$ is a vector of nonnegative entries on the simplex
$$\boldsymbol{\Delta}^{d-1}=\left\{(u_1,\ldots,u_d)': u_k\ge0\;\forall k, \sum_{k=1}^du_k=c\right\},$$
where we take $c=1$ without loss of generality. The questions that motivate these data though are rarely about the proportions themselves. They concern which taxa or functional groups occur in a given sample, which groups respond together to shared environmental drivers (e.g., temperature, nutrients, light, depth), and how those mechanisms shift across space and time. Simplex closure complicates the path from observed proportions to the underlying biological or ecological process; the unit-sum constraint forces components to compete and can leave apparent compositional correlations weak or even reversed in sign relative to the latent abundance-scale association.

Two recurring features of community composition data make this mismatch concrete. The first is exact zeros. They are common, and their meanings are not interchangeable; a zero may signal a structural absence, a habitat-driven exclusion, a dispersal limitation, or a detection failure below an effective threshold. Replacing such zeros with small positive values is computationally convenient, yet it conflates presence with relative dominance and can manufacture dependence that is not there. The second feature is shared latent drivers; groups of taxa or functional categories often respond positively to the same environmental gradient even as closure forces their observed proportions to compete. A model that captures only one of these features leaves the other to contaminate the inference.

Existing approaches handle only part of this problem. Dirichlet-type models honor compositional geometry but, in their basic form, tie the covariance to the mean and a single concentration parameter. The generalized Dirichlet extension uses a stick-breaking construction that admits limited negative correlations \citep{connor1969concepts}, and the Dirichlet-tree multinomial  decomposes the multinomial likelihood along a phylogenetic tree and parameterizes its $K\times K$ covariance with $O(K)$ parameters \citep{dennis1991hyper, wang2017dirichlet}. Such extensions enrich the dependence structure but keep a parametric form that still cannot represent the arbitrary positive correlations induced by shared latent drivers. Logistic-normal models \citep{aitchison1982statistical,egozcue2003isometric} are more flexible because a log-ratio map transforms compositions to Euclidean space, where a latent Gaussian flexibly models covariance. However, log-ratio transformations are undefined at zeros and, therefore, awkward under sparsity. A separate line of work instead models structural zeros explicitly through point masses on the simplex. \cite{leininger_spatial_2013} introduced a power-scaling approach with multivariate latent variables that govern zero/nonzero entries in a spatial regression for compositional data with many zeros. \cite{butler2008latent} treated zeros geometrically, projecting a latent multivariate normal vector orthogonally onto the simplex boundary, and \cite{tsagris2022modelling} refined this by projecting along the line from each point to the simplex centroid and fitting a zero-censored multivariate normal model in $\alpha$-transformed coordinates. Such constructions do accommodate structural sparsity. However, the projection-based methods rest on \textit{ad hoc} geometric choices for assigning boundary mass, while the power-scaling and $\alpha$-transformations obscure the directional geometry of the original composition; these approaches fail to capture the positive latent dependence that arises when components share environmental drivers.

A parallel literature in microbial community analysis considers exact-zero handling and latent dependence in factor-augmented multinomial, Dirichlet, or logistic-normal models. \citet{datta2016bayesian} developed a quasi-sparse count model that supplies the foundational shrinkage framework for subsequent compositional applications. \citet{tang2019zero} developed a zero-inflated generalized Dirichlet-multinomial regression. \citet{ren2020bayesian} developed Bayesian mixed-effects models for zero-inflated compositions in which dependence across compositions is induced through latent factors with adaptively learned dimensions. \citet{xu2021zero} introduced zero-inflated Poisson factor analysis, and \citet{zeng_zero-inflated_2023} developed a zero-inflated probabilistic PCA logistic-normal multinomial framework. \citet{koslovsky_bayesian_2023} introduced a Bayesian zero-inflated Dirichlet-multinomial regression for compositional count data, and \citet{datta2024bayesian} developed horseshoe-shrinkage priors for compositional regression with application to oral-microbiome compositions. A related thread targets microbial covariation networks rather than factor structure; \citet{ha2020compositional} introduced a multivariate hurdle graphical model that jointly infers conditional dependencies among continuous abundances and presence/absence indicators, and \citet{xu2026bayesian} developed a covariance regression in which the network rewires with covariates through a sparse-plus-low-rank decomposition. These approaches parameterize dependence through Dirichlet-type covariances or log-ratio specifications that latent factors only partially relax, and the dependence they infer lives on the count or log-ratio scale rather than the abundance scale on which shared environmental responses most directly act.

Projected Gaussian constructions offer a different geometric route, one that preserves directional structure by normalizing a latent Gaussian vector. An elementwise square root transformation maps from the simplex to the nonnegative orthant of the unit hypersphere, where directional distributions may be used \citep{stephens1982use}. \cite{paine_elliptically_2018} introduced the elliptically symmetric angular Gaussian (ESAG) distribution, which pairs an analytically tractable normalizing constant with anisotropic latent covariance across the entire hypersphere. However, for square-root transformed compositions, the relevant space is the nonnegative orthant, and conditioning a projected Gaussian there produces a clipped likelihood whose normalizing constant is a Gaussian orthant probability (i.e., the probability that a Gaussian vector with the current mean and covariance lands in $\R^d_+$). This constant is parameter-dependent, varies across observations and across MCMC iterations in Bayesian implementations, and grows computationally prohibitive as $d$ increases. Standard orthant-probability algorithms compound the difficulty; the Geweke-Hajivassiliou-Keane estimator, for instance, has relative variance that diverges exponentially in $d$ \citep{ridgway2016computation}, which leaves the truncated likelihood unstable at moderate to high dimensions. \cite{schwob2025spatial} confronted this barrier directly by approximating the orthant probability via Monte Carlo at every parameter value that the sampler visits, but the construction remains feasible only at low dimensions (e.g., $d< 100$) and does not separately accommodate exact zeros.

This work aims to keep the latent Gaussian interpretation of projected Gaussian models while accommodating exact zeros explicitly, allowing flexible positive and negative latent dependence, and eliminating the orthant normalizing constant altogether. The key idea is a change of perspective; a composition is taken as the observed direction of a latent abundance vector rather than as a point in the simplex, with magnitude unobserved and an active set naming the components that are present. The manuscript provides four contributions: (i) we propose the Active-set Ray-based Compositional (ARC) model for sparse high-dimensional compositional data and derive its likelihood with the radius treated as an auxiliary latent variable; (ii) we then extend the model to structural zeros through an active-set formulation that splits the presence/absence process from the positive subcomposition such that each is specified by its own likelihood factor with its own covariates, latent factors, and priors; (iii) to keep inference tractable, we introduce a low-rank factor covariance with a probit active-set model, a pairing that yields conditionally Gaussian updates and scales to high-dimensional sparse compositions; (iv) finally, we show that the active-set process and the latent positive dependence can be recovered jointly on sparse high-dimensional compositional data.

The paper proceeds as follows. Section \ref{sec:comp-geo} develops compositional geometry and the hyperspheric embedding. Section \ref{sec:zeros} introduces the active-set process and discusses what a structural zero means, and Section \ref{sec:ray-model} specifies the ARC model, derives its likelihood, and builds the regression framework. The Bayesian implementation follows in Section \ref{sec:implementation}. Section \ref{sec:simulation-study} reports a simulation study, and 
Section \ref{sec:discussion} closes with extensions to spatial, spatio-temporal, and detection-error settings and discusses implications for microbial, ecological, and environmental applications.

\section{Compositional Geometry and the Positive Hypersphere}
\label{sec:comp-geo}

Two conventional modeling routes respect compositional geometry. One transforms the composition into Euclidean space \citep[e.g., log-ratio modeling;][]{aitchison1982statistical, egozcue2003isometric}, which removes the closure constraint and yields a coordinate system on which Gaussian models are well-defined. However, log-ratio transformations are undefined when components are zero-valued, which is common in microbial and ecological data. Several remedies soften the positivity constraint. Multiplicative and Bayesian-multiplicative replacement methods substitute model-based small values for zeros before transforming \citep{martin2003dealing,martin2015bayesian}, and pivot-coordinate constructions only require that one chosen pivot component be strictly positive \citep{hron2017weighted}. Although these approaches impose weaker constraints, they do not remove the underlying reliance on positivity. An alternative route focuses directly on the simplex and considers Dirichlet-type distributions, which admit exact zeros only as a measure-zero boundary limit and remain parametrically restricted in their dependence \citep{connor1969concepts, dennis1991hyper, wang2017dirichlet, datta2024bayesian}.

\begin{figure}[h]
\centering
\includegraphics[width=\linewidth]{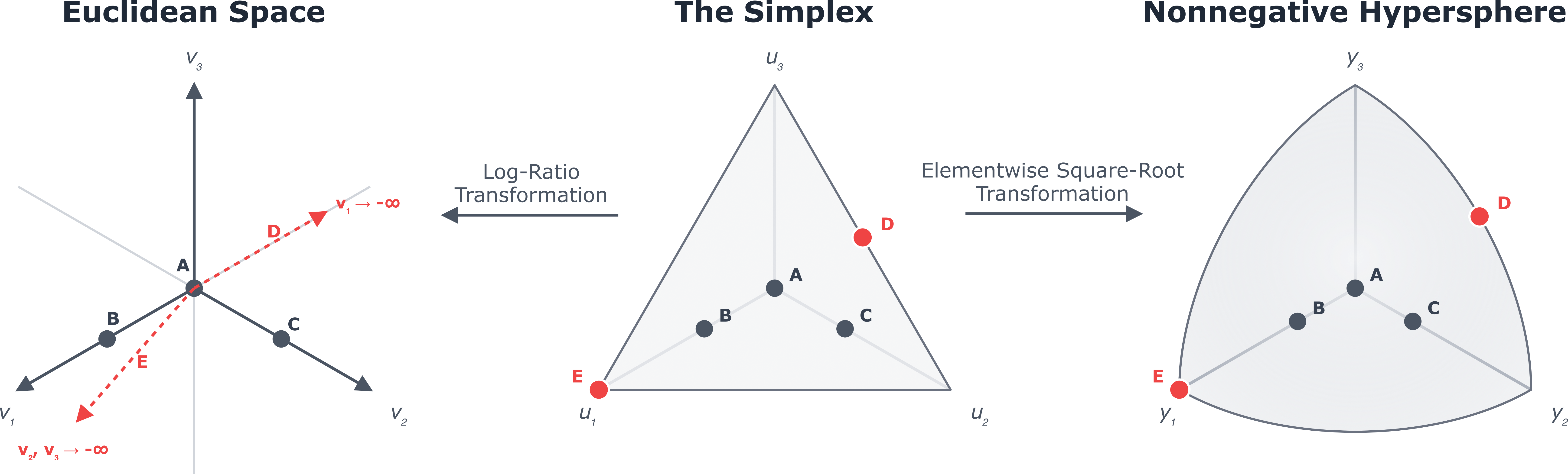}
\caption{Common modeling spaces for three-dimensional compositional data with five observations. \emph{Center}: The simplex $\Delta^2$, which is the natural domain of compositional data. \emph{Left}: Euclidean $\R^3$, into which the simplex maps via log-ratio transformations; observations $D$ and $E$ are not properly transformed because they contain zeros, and the fitted covariance is singular along $\bfone$. \emph{Right}: The nonnegative orthant of the unit hypersphere $\Sph^2_+$, reached via the elementwise square root transformation $\bfy = \sqrt{\bfu}$; exact zeros lie naturally on geodesic arc boundaries.}
\label{fig:geometry}
\end{figure}

We take a third route and embed the composition in the nonnegative orthant of the unit hypersphere. An elementwise square root transforms a composition $\bfu$ to a directional point $\bfy=\sqrt{\bfu}\in \Sph^{d-1}_{0,+}$, where
\[
    \Sph^{d-1}_{0,+}
    =
    \left\{\bfy\in\R^d : \|\bfy\|=1,\ y_k\geq 0\ \text{for all } k\right\}
\]
denotes the nonnegative orthant of the unit hypersphere and $\|\cdot\|$ denotes the Euclidean norm. Beyond handling each difficulty just described, this embedding offers two advantages that the simplex does not. The first concerns the constraint itself because the unit-sum constraint $\bfone'\bfu_i = 1$ becomes a unit-norm constraint
\[
    \|\bfy\|^2
    =
    \sum_{k=1}^d y_{k}^2
    =
    \sum_{k=1}^d u_{k}
    = 1,
\]
which is the defining property of the unit hypersphere \citep{mardia2009directional,scealy_regression_2011,paine_elliptically_2018}. Working on the hypersphere is advantageous because directional distributions may be built from latent Euclidean drivers by projection (i.e., $\bfy = \bfx / \|\bfx\|$ for $\bfx \in \R^d$), which provides a latent Gaussian structure with flexible covariance modeling. The embedding is also faithful at the boundary. Because the square root preserves zeros exactly, $u_{k} = 0 \Longleftrightarrow y_{k} = 0$ $\forall k$, exact zeros survive as boundary points on the nonnegative hypersphere and do not need regularization, replacement, or separate transformation. In contrast, log-ratio maps are undefined at the boundary and simplex-based constructions require strictly positive compositions for the inner product to be well-defined. The hyperspheric embedding accommodates exact zeros without replacement or transformation and retains a latent Gaussian representation that is well-suited for flexible covariance modeling.

\section{Structural Zeros and the Active Set}
\label{sec:zeros}

Conventional compositional analyses treat zeros as numerical inconveniences. We instead make the zeros part of the model by defining the active set
\[
    A_i = \{k : y_{ik} > 0\} = \{k : u_{ik} > 0\},
\]
where $m_i = |A_i|$ is the number of active components in $\bfu_i$. The components excluded from $A_i$ are taken to be structural zeros for observation $i$. Conditional on $A_i$, the positive subcomposition $\bfy_{i}^+$ lies on a lower-dimensional positive hypersphere $\Sph^{m_i-1}_{+}$ such that all entries are strictly positive and $\sum_{k\in A_i} y_{ik}^2 = 1$.

The active set induces a factorization of the likelihood $[\bfy_i\mid \cdot]$:
\begin{equation}\label{eq:factored-likelihood}
    [\bfy_i\mid\cdot] = [\bfy_{i}^+\mid A_i,\cdot]\,[A_i\mid\cdot],
\end{equation}
where $[\cdot\mid\cdot]$ denotes the conditional probability density or mass function \citep{gelfand1990sampling}.
Such a factorization is principled in microbial and ecological applications, where presence and dominance are governed by related but distinct mechanisms. For example, presence may depend on dispersal, habitat suitability, seasonality, detection thresholds, or structural unavailability, whereas dominance among the present components may depend on competition, shared environmental gradients, or latent community structure. A single continuous density on the simplex blurs such a distinction because one set of density parameters must simultaneously encode whether a component is present and, given presence, how strongly it dominates; these are separable mechanisms that may not respond to the same drivers.

We consider two specifications for the active-set process $[A_i\mid \cdot]$. We start with conditionally independent inclusion indicators with covariate-driven probabilities:
\[
    Z_{ik} = \mathbf{1}(k\in A_i)\sim\operatorname{Bernoulli}(\rho_{ik}),
    \qquad
    \logit(\rho_{ik}) = \bfw_i'\bfalpha_k + \bfh_i'\bfgamma_k,
\]
where $\bfw_i$ comprises covariates such as spatial coordinates, time, environmental drivers, or sampling-effort summaries, $\bfalpha_k$ are the occurrence covariate coefficients, $\bfh_i$ comprises basis functions evaluated at the observation location, and $\bfgamma_k$ are the occurrence basis coefficients.
Such a specification is parsimonious and scalable but does not allow for the residual co-occurrence common in ecological and microbial data. To capture correlated occurrence, we add latent factors:
\[
    \logit(\rho_{ik})
    =
    \bfw_i'\bfalpha_k + \bfh_i'\bfgamma_k + \bfomega_k'\bm{o}_i,
    \quad
    \bm{o}_i\sim \text{N}_{q_o}(\bfzero,\bfI_{q_o}), \quad i=1,\dots,n, \; k=1,\dots,d,
\]
where $\bm{o}_i \in \R^{q_o}$ is the occurrence factor for observation $i$, $\bfomega_k \in \R^{q_o}$ is the occurrence loading for component $k$, and $q_o$ is the rank of the occurrence factor model. 
Components with similar occurrence loadings tend to appear together, which captures co-occurrence and mutual exclusion without enumerating all $2^d$ active sets. In our simulation study, we adopt a probit analogue of the second specification to facilitate Albert-Chib augmentation for high-dimensional compositional data \citep{albert1993bayesian}.

We model the active set as a discrete presence/absence process rather than through a continuous prior that shrinks small components toward zero, which is often preferred in sparse problems \citep{stephens2017false}. The case for the continuous approach rests on priors over latent effects, where a point mass at zero idealizes an underlying continuous, unimodal effect distribution. Compositional zeros are different because they are a feature of the observed data. A compositional zero is exact ($u_{ik} = 0$), and the square-root transformation preserves it as an exact boundary point ($y_{ik} = 0$), so the active set models an observed, exact event rather than a shrinkage idealization.

The factorization in \eqref{eq:factored-likelihood} also clarifies what a particular zero represents. In community composition data, an observed zero may reflect a genuine structural absence (i.e., a taxon ecologically excluded from the location, an unsuitable habitat, or a dispersal limitation), or it may reflect a detection failure in which the taxon is present below an effective sampling threshold and is therefore recorded as absent. The two interpretations carry different scientific implications. Structural absences inform which components belong to a community and which environmental gradients exclude them, whereas detection failures inform sampling design and effort. A model that treats all zeros identically conflates these meanings, and it may attribute a detection-driven gap to an ecological process or, conversely, dismiss a genuine absence as a sampling error.

By relocating the zero process from a density-level nuisance to a modeled event, the active-set layer accommodates both types of zeros. When a zero marks a structural absence, $Z_{ik}=0$ and inference proceeds directly. A detection failure is handled differently; $Z_{ik}=1$ is reinterpreted as an observed presence of a latent true active set, and the observation model introduces component-specific false-negative probabilities that depend on covariates such as sampling effort, depth, or sequencing depth in microbial applications. Applying the ray-based positive subcomposition model $[\bfy_i^+\mid A_i,\cdot]$ to the latent positive components allows the detection layer to absorb sampling variation. Thus, the two interpretations share the same positive subcomposition mechanism yet differ only in the active-set layer. As a result, structural and sampling zeros receive a unified treatment within a single model.

\section{The Active-set Ray-based Compositional (ARC) Model}
\label{sec:ray-model}

Conditional on $A_i$, we model the positive subcomposition $\bfy_i^+$ through a latent Gaussian density evaluated along the ray generated by $\bfy_i^+$. We introduce a latent radius $r_i > 0$ and the latent active vector
\[ \bfx_i^+ = r_i \, \bfy_i^+, \]
which lies in the positive orthant of the $m_i$-dimensional active subspace because $\bfy_i^+$ is elementwise positive and $r_i > 0$. Because $r_i$ is unobserved, $\bfx_i^+$ is latent rather than observed; the data inform $\bfx_i^+$ only through its direction $\bfy_i^+$. We assign the base distribution
\[ \bfx_i^+ \sim \text{N}_{m_i}(\bfmu_i^+, \bfV_{A_iA_i}), \]
where $\bfmu_i^+\in\R^{m_i}$ denotes the active entries of the latent mean $\bfmu_i\in\R^d$ and $\bfV_{A_iA_i}\in\R^{m_i\times m_i}$ denotes the active submatrix of the covariance $\bfV \in \R^{d \times d}$. The unrestricted normal places mass throughout $\R^{m_i}$, but the observation enters only through the positive rays of the active subspace such that every realization $\bfx_i^+ = r_i \bfy_i^+$ lies in the positive orthant by construction.
The full-dimensional latent mean $\bfmu_i$ and covariance $\bfV$ describe the abundance-scale Gaussian vector before simplex closure, and $r_i$ is the unobserved magnitude that is lost when the data are reduced to relative compositions. 
The radius is identifiable up to the parameterization of the base distribution. In particular, the likelihood is unchanged under the joint rescaling $\{r_i, \bfbeta_k, \bfdelta_k, \bflambda_k, \sigma_k\} \mapsto \{sr_i, s\bfbeta_k, s\bfdelta_k, s\bflambda_k, s\sigma_k\}$ for any $s > 0$; we address this in Section \ref{sec:implementation}.

The proposed construction is closely related to the projected normal distribution, in which a Gaussian density on $\R^d$ generates a directional density on the unit hypersphere through normalization \citep{wang2013directional, hernandez2017general}. The ARC model specializes this construction to the positive rays of the active subspace such that the unrestricted Gaussian base distribution is interpreted not generatively over $\R^{m_i}$ but only along those rays.

The radius augmentation itself is not new. \citet{nunez2005bayesian} introduced latent length variables for the projected normal distribution in the circular case, and \citet{hernandez2017general} extended this to arbitrary dimension under a reparameterization that yields closed-form Gaussian full conditionals and a slice sampler for the radii. We adopt the same augmentation, though its role here differs. In the projected normal literature, the latent length aids posterior simulation under an unconstrained projected normal likelihood, whereas in the ARC model the latent radius ensures support on the positive hypersphere and removes the need to evaluate the orthant probability, which is what renders the construction tractable in high dimensions.

To obtain the density of the observed direction $\bfy_i^+$, we change variables from the latent active vector $\bfx_i^+ \in \R^{m_i}$ to polar coordinates, namely the radius $r_i = \|\bfx_i^+\|$ and the direction $\bfy_i^+ = \bfx_i^+ / \|\bfx_i^+\|$; the direction lies in $\Sph^{m_i-1}_{0,+}$ because $\bfx_i^+$ is elementwise positive. The Euclidean volume element (i.e., the Lebesgue measure $d\bfx_i^+$ on $\R^{m_i}$) then factorizes as
\[
    d\bfx_i^+ = r_i^{m_i-1}\,dr_i\,d\sigma(\bfy_i^+),
\]
where $d\sigma$ is the surface measure on the unit sphere $\Sph^{m_i-1}$ and $r_i^{m_i-1}$ is the Jacobian. Applying this factorization to the Gaussian density of $\bfx_i^+$, the joint density of $\{r_i,\bfy_{i}^+\}$ conditional on $A_i$ is
\begin{align}
    [r_i,\bfy_{i}^+\mid A_i,\bfmu_i,\bfV]
    &=
    (2\pi)^{-m_i/2}|\bfV_{A_iA_i}|^{-1/2}\,r_i^{m_i-1} \notag\\
    &\quad\times
    \exp\!\left\{-\tfrac{1}{2}
    (r_i\bfy_{i}^+-\bfmu_{i}^+)'
    \bfV_{A_iA_i}^{-1}
    (r_i\bfy_{i}^+-\bfmu_{i}^+)
    \right\},
    \qquad r_i>0.
    \label{eq:ray_joint}
\end{align}
Integrating out $r_i$ gives the marginal density on the positive orthant of the unit hypersphere:
\begin{equation}
    [\bfy_{i}^+\mid A_i,\bfmu_i,\bfV]
    =
    \int_0^\infty
    [r_i,\bfy_{i}^+\mid A_i,\bfmu_i,\bfV]\,dr_i.
    \label{eq:ray_marginal}
\end{equation}
In \eqref{eq:ray_joint} and \eqref{eq:ray_marginal}, the quantities $\{\bfx_i^+, \bfy_i^+,\bfmu_i^+\}$ are $m_i$-vectors indexed by the active components $A_i$; the full-dimensional latent means $\{\bfmu_i\}$ and the covariance $\bfV$ enter only through their active subvectors $\{\bfmu_i^+\}$ and the active submatrix $\bfV_{A_i A_i}$. Because the radius is retained as an auxiliary variable, \eqref{eq:ray_joint} suffices for posterior computation.
Equation \eqref{eq:ray_joint} is written as an equality rather than a proportionality because the ARC construction does not normalize over the positive orthant. The Gaussian density serves as a likelihood kernel evaluated along observed positive rays, not as a generative distribution whose support coincides with $\Sph^{m_i-1}_{0,+}$. A truncated directional model would instead require the orthant probability $P_{A_i}(\bfmu_i, \bfV) = P(\bfx_i^+ \in \R_+^{m_i})$, which is the computationally prohibitive term that ARC avoids \citep{schwob2025spatial}. Notably, the data $\bfy_i^+ \in \Sph^{m_i-1}_{0,+}$ enter the likelihood along positive rays of the active subspace; the Gaussian behavior over the rest of $\R^{m_i}$ never enters inference.

The ARC construction provides two benefits, and the active set $\{A_i\}$ is central to both. The first concerns scale. ARC infers dependence on the latent abundance scale rather than on the closed compositional scale; this matters because the unit-sum constraint induces structural negative dependence among the observed proportions even when the underlying components are positively associated through shared environmental drivers. A model fit directly on the simplex therefore conflates the latent biological signal with the distortion that closure imposes. Parameterizing dependence on the abundance-scale latent vector $\bfx_i^+$ separates the two such that the recovered covariance $\bfV$ reflects biological co-response rather than the algebra of summing to one. The second benefit is computational. The construction in \eqref{eq:ray_joint} contains no orthant normalizing constant; a truncated directional model, by contrast, would condition a projected Gaussian on $\bfy\in\Sph^{d-1}_{0,+}$ and incur the parameter-dependent orthant probability $P_{\bfmu_i,\bfV}(\bfY_i\in\Sph^{d-1}_{0,+})$, whose evaluation is computationally prohibitive. Here, $\bfy_{i}^+$ is already positive and $r_i > 0$, so conditioning on $A_i$ places $\bfx_i^+ = r_i\bfy_{i}^+$ in the positive orthant by construction. Thus, the active set plays two roles at once: a scientific one, in separating presence from dominance, and a computational one, in keeping the ARC likelihood tractable at high dimension while preserving the latent Gaussian interpretation.

The proposed likelihood-kernel approach also suggests where the ARC model is appropriate. It is best suited for inference on observed compositions, not for unconditional simulation.
Given an observed positive direction $\bfy_i^+ \in \Sph^{m_i-1}_{0,+}$ and active set $A_i$, the expression in \eqref{eq:ray_joint} is the proper joint density of the polar decomposition of $\bfx_i^+$ evaluated at the observed direction, and posterior inference proceeds along positive rays.
An unconditional draw $\bfx_i^+ \sim \text{N}_{m_i}(\bfmu_i^+, \bfV_{A_iA_i})$ followed by normalization is another matter; it can yield directions outside $\Sph^{m_i-1}_{0,+}$ because the unrestricted Gaussian places mass throughout $\R^{m_i}$, which is exactly the limitation that truncated directional approaches handle through orthant conditioning. We trade unconditional generativity for tractable inference, which is most appropriate when the goal is to recover latent structure from observed compositions rather than to simulate new ones.

\subsection{Regression, Structured Variation, and Scalable Covariance}

Compositional analyses are frequently concerned with how environmental factors shape composition, which we examine by regressing the latent mean $\bfmu_i$ on covariates $\bfw_i$:
\begin{equation}\label{eq:mean-spec}
    \bfmu_i = \bfB\bfw_i + \bfDelta\bfh_i,
\end{equation}
where $\bfB\in\R^{d\times p}$ is a coefficient matrix whose $k$th row records how the latent abundance tendency of component $k$ shifts with the covariates prior to simplex closure, $\bfDelta\in\R^{d\times L}$ carries each component's basis effects in its corresponding row, and $\bfh_i\in\R^L$ holds the values of radial basis functions evaluated at the observation location. Modeling the mean in such a way brings the data-generating process closer to ecological and microbial settings, in which composition is shaped jointly by environmental gradients and by residual structure that the covariates do not capture. We sketch a spatial extension to \eqref{eq:mean-spec} in Section \ref{sec:discussion}.

The covariance matrix $\bfV$ captures the dependence among the active components and is the principal target of inference. Because a dense $d\times d$ covariance is infeasible at large dimension, we adopt the low-rank factor form
\begin{equation}\label{eq:lowrank}
    \bfV = \bfLambda\bfLambda' + \bfD,
\end{equation}
which is built from a loading matrix $\bfLambda\in\R^{d\times q}$ ($q\ll d$) and a diagonal matrix $\bfD = \diag(\sigma_1^2,\ldots,\sigma_d^2)$ that gathers the component-specific residual variances $\{\sigma_k^2\}$ (i.e., the uniquenesses of factor analysis). Equivalently,
\[
    \bfx_i = \bfmu_i + \bfLambda\bff_i + \bfepsilon_i,
    \qquad
    \bff_i\sim \text{N}_q(\bfzero,\bfI_q),
    \qquad
    \bfepsilon_i\sim \text{N}_d(\bfzero,\bfD),
\]
which reduces the covariance from $O(d^2)$ to $O(dq + d)$ free parameters and provides interpretable latent factors $\{\bff_i\}$. For a given observation, only the active submatrix
\[
    \bfV_{A_iA_i} = \bfLambda_{A_i}\bfLambda_{A_i}' + \bfD_{A_i}
\]
is needed, which is convenient when the active sets are sparse. Whenever $m_i\ll d$, the likelihood involves only the active components such that the factor-augmented sampler can replace repeated dense inversions of $\bfV_{A_iA_i}$ with conditionally Gaussian updates over the active entries, and the effective per-observation dimension becomes the number of nonzero components $m_i$ rather than $d$.

The columns of $\bfLambda$ stand for shared gradients or unobserved drivers that tie the components together. In microbial and ecological applications, components may behave as a functional group (e.g., a shared metabolic guild or a common response to an unobserved environmental gradient) that rises and falls together; that is, similar loadings covary.
Thus, $\bfV$ is not a block of pairwise correlations but a small number of community-level drivers, which is precisely what ecological and microbial studies of shared environmental response investigate.
In addition to interpretability, another advantage is identifiability.
The factor model inherits the usual rotational nonidentifiability of $\bfLambda\bfLambda'$, but the quantities of interest $\bfV = \bfLambda\bfLambda' + \bfD$ and $\bfR = \diag(\bfV)^{-1/2}\bfV\diag(\bfV)^{-1/2}$ remain fully identifiable \citep{anderson1956statistical, geweke1996measuring}.

\section{Implementation}
\label{sec:implementation}

We implement the ARC framework as a Bayesian hierarchical model fit by MCMC with two augmentations doing most of the work. The first is the latent radius $r_i > 0$, which turns the observed spherical direction into an augmented Gaussian vector along the observed ray. The second supplies latent factors for the covariance model together with Albert-Chib latent variables for the probit active-set model. Conditional on these augmented quantities, most of the model collapses into a collection of efficient Gaussian regression problems whose missing entries are induced by the active sets.

\subsection{Radius Update}

The radius density for observation $i$ follows from \eqref{eq:ray_joint}. Its full conditional is
\[
    [r_i\mid \bfy_{i}^+, A_i, \bfmu_i,\bfV]
    \propto
    r_i^{m_i-1}
    \exp\!\left\{-\tfrac{1}{2}a_i r_i^2 + b_i r_i\right\},
    \qquad r_i > 0,
\]
with
\[
    a_i = (\bfy_{i}^+)'\bfV_{A_iA_i}^{-1}\bfy_{i}^+,
    \qquad
    b_i = (\bfy_{i}^+)'\bfV_{A_iA_i}^{-1}\bfmu_{i}^+.
\]
This update is one-dimensional regardless of $d$ and never invokes an orthant probability, so its cost scales with $m_i$ through the construction of $a_i$ and $b_i$ rather than with $d$. We use a random-walk Metropolis update on $\log r_i$ in the simulation and case study.

Because the likelihood is scale-equivariant under the joint rescaling, the chain mixes slowly along the unidentified scale direction. To stabilize posterior summaries of the scale-dependent quantities $\{r_i, \bfbeta_k, \bfdelta_k, \bflambda_k, \sigma_k\}$ without changing the model or the priors, we rescale each saved draw \textit{post hoc} such that $\frac{1}{n}\sum_i r_i$ equals its initial value $\sqrt{\bar m}$, where $\bar m = \frac{1}{n}\sum_i m_i$. The rescaling simply selects one representative from each likelihood-equivalent orbit. Therefore, the identifiable summaries $\bfR$ and $\bm{\rho}$ are invariant, and inference on the active-set parameters $\{\bfalpha_k, \bfgamma_k, \bfomega_k\}$ is unaffected.

\subsection{Positive Subcomposition Layer}

Conditional on $r_i$, we represent the positive subcomposition as a Gaussian factor model with missing entries:
\begin{equation}\label{eq:factor_rep}
    r_i y_{ik}
    =
    \bfw_i'\bfbeta_k
    + \bfh_i'\bfdelta_k
    + \bff_i'\bflambda_k
    + \epsilon_{ik},
    \qquad
    \epsilon_{ik}\sim \text{N}(0,\sigma_k^2), \qquad k \in A_i,
\end{equation}
where $\bfbeta_k = [\bfB]_{k,\cdot}^\prime \in \R^p$ is the vector of covariate effects for component $k$, $\bfdelta_k = [\bfDelta]_{k,\cdot}^\prime \in \R^L$ is its vector of basis effects, and $\bflambda_k = [\bfLambda]_{k,\cdot}^\prime \in \R^q$ is its factor-loading vector. This representation yields conditionally Gaussian updates for the factors $\{\bff_i\}$:
\[
    [\bff_i\mid\cdot]
    \sim
    \text{N}_q\!\left(\bfQ_{f_i}^{-1}\bfh_{f_i},\,\bfQ_{f_i}^{-1}\right),
\]
where
\[
    \bfQ_{f_i}
    = \bfI_q + \sum_{k\in A_i}\frac{\bflambda_k\bflambda_k'}{\sigma_k^2},
    \qquad
    \bfh_{f_i}
    = \sum_{k\in A_i}\frac{\bflambda_k\,e_{ik}^{(-f)}}{\sigma_k^2},
\]
and $e_{ik}^{(-f)} = r_i y_{ik} - \bfw_i'\bfbeta_k - \bfh_i'\bfdelta_k$ collects the residuals excluding the current factor term.

The component-specific regression and loading vectors are likewise updated from Gaussian full conditionals once the factors $\{\bff_i\}$ are fixed. We let $I_k = \{i : k\in A_i\}$ be the set of observations in which component $k$ is active. For $i\in I_k$, we set $x_{ik}^{\ast} = r_i y_{ik}$ such that
\begin{equation}\label{eq:conj_full}
    x_{ik}^{\ast}
    = \bfw_i'\bfbeta_k + \bfh_i'\bfdelta_k + \bff_i'\bflambda_k + \epsilon_{ik},
\end{equation}
which, given the factors $\{\bff_i\}$, radii $\{r_i\}$, and uniquenesses $\{\sigma_k^2\}$, is a standard Gaussian linear model. Under Gaussian priors on $\{\bfbeta_k, \bfdelta_k, \bflambda_k\}$, each parameter vector has a conjugate full conditional. Then, the uniquenesses $\{\sigma_k^2\}$ are drawn from an inverse-gamma full conditional built from the residual sum of squares over $I_k$ under a conjugate inverse-gamma prior.

\subsection{Active-Set Layer}

The active-set process is modeled separately through a probit factor specification, which gives a clean conditionally Gaussian sampler. With $Z_{ik} = \mathbf{1}(k \in A_i)$ indicating whether component $k$ is active in composition $i$, the probit model thresholds a latent Gaussian variable
\[
    Z_{ik}^* = \bfw_i'\bfalpha_k + \bfh_i'\bfgamma_k + \bfomega_k'\bm{o}_i + e_{ik}, \qquad e_{ik} \sim \text{N}(0,1), \qquad Z_{ik} = \mathbf{1}(Z_{ik}^* > 0),
\]
where $\bfalpha_k$ and $\bfgamma_k$ are component-specific covariate and basis coefficients for occurrence, $\bm{o}_i \in \R^{q_o}$ is the occurrence factor for observation $i$, and $\bfomega_k \in \R^{q_o}$ is the occurrence loading for component $k$. Albert-Chib augmentation draws $Z_{ik}^{\ast}$ from a normal distribution truncated above or below zero according to $Z_{ik}$ \citep{albert1993bayesian}, and, conditional on $Z_{ik}^{\ast}$, the updates for $\{\bfalpha_k, \bfgamma_k, \bm{o}_i, \bfomega_k\}$ are conjugate Gaussian.

\subsection{Computational Costs}

Algorithm \ref{alg:sampler} summarizes the full sampler, which we implement in Julia for efficiency \citep{bezanson2017julia}. Its per-iteration cost is governed by the active-set size and the factor rank, not by $d$. We let $M = \sum_i m_i \leq nd$ denote the total number of active entries, with $M \ll nd$ under sparsity. The probit latent updates cost $O(nd)$, one truncated-normal draw per entry. Each component-specific update for $\{\bfalpha_k, \bfgamma_k, \bfbeta_k, \bfdelta_k, \bflambda_k, \bfomega_k\}$ solves a Gaussian system of dimension $O(p + L + q)$, which accumulates to $O\!\left(d(p+L+q)^3 + (p+L+q)^2 M\right)$ across active entries. The factor updates for $\bff_i$ and $\bm{o}_i$ require $q$- and $q_o$-dimensional Gaussian solves costing $O(q^3 n + q^2 M)$ and $O(q_o^3 n + q_o^2 M)$, respectively. Radius updates contribute quadratic forms in $\bfV_{A_i A_i}^{-1}$ at $O\!\left(\sum_i m_i^2\right)$, and the uniqueness updates are $O(M)$. Therefore, the dominant term is $O\!\left(d(p+L+q)^3 + (p+L+q)^2 M + q^3 n\right)$ per iteration, which is linear in $d$ for fixed factor rank and sparse active sets.

\begin{algorithm}
\caption{ARC MCMC sampler}
\label{alg:sampler}
\begin{algorithmic}[1]
\State Initialize $\{Z_{ik}^*\}$, $\{r_i\}$, $\{\bm{o}_i\}$, $\{\bff_i\}$, and component-specific parameters.
\For{iteration $t = 1, \ldots, T$}
    \State Update radii $\{r_i\}$ from their one-dimensional full conditionals.
    \State Update positive subcomposition factors $\{\bff_i\}$ from Gaussian full conditionals.
    \State Update positive subcomposition regression coefficients $\{\bfbeta_k\}$, basis coefficients $\{\bfdelta_k\}$, factor loadings $\{\bflambda_k\}$, and uniquenesses $\{\sigma_k^2\}$ from Gaussian and inverse-gamma full conditionals.
    \State Update probit latent variables $\{Z_{ik}^*\}$ for the active-set model.
    \State Update active-set regression coefficients $\{\bfalpha_k\}$, basis coefficients $\{\bfgamma_k\}$, occurrence factors $\{\bm{o}_i\}$, and occurrence loadings $\{\bfomega_k\}$ from Gaussian full conditionals.
\EndFor
\end{algorithmic}
\end{algorithm}

In contrast, truncated directional approaches must evaluate a Gaussian orthant probability $P_{\bfmu_i, \bfV}(\bfY_i \in \Sph^{d-1}_{0,+})$ at every observation and every iteration \citep{schwob2025spatial}. 
If $M_{\text{MC}}$ is the number of Monte Carlo draws used per evaluation, this estimation costs $O(M_{\text{MC}} n d^2)$ per iteration, and the relative variance of standard orthant-probability estimators grows exponentially with $d$ \citep{ridgway2016computation}. Such an approach is feasible only at low dimensions. Dirichlet-multinomial constructions sidestep orthant probabilities and scale as $O(nd)$ per iteration through digamma or polynomial evaluations, but they impose restrictive dependence structures that cannot represent arbitrary positive latent correlations \citep{tang2019zero, koslovsky_bayesian_2023, datta2024bayesian}.
Logistic-normal multinomial constructions with factor-augmented covariance scale comparably to the ARC sampler at $O(d(p+q)^3 + nd)$ per iteration, yet they require log-ratio transformation or hurdle-component zero handling that ties the inferred dependence to the count or log-ratio scale rather than the abundance scale of interest \citep{zeng_zero-inflated_2023}.

Two further features help the framework scale. For very large $d$, shrinkage priors on $\bfLambda$, sparse factor models, or adaptive rank selection can be incorporated without altering the construction \citep{bhattacharya2011sparse}. For very large $n$, the updates decouple across $i$ (for $\bff_i$, $\bm{o}_i$, $r_i$) and across $k$ (for $\bfbeta_k$, $\bflambda_k$, $\bfomega_k$, $\sigma_k^2$), so observation-wise and component-wise updates parallelize naturally.

When a component is active in few observations relative to its coefficient dimension, the expensive $O((p+L+q)^3)$ Cholesky solve in the component-specific update can be avoided. The sampler of \citet{bhattacharya2016fast} draws from the relevant conditionally Gaussian posterior in $O\!\left(|I_k|^2 (p+L+q)\right)$ operations, which is linear rather than cubic in the coefficient dimension and advantageous whenever $|I_k|$ is small (i.e., when a component is active in far fewer observations than its coefficient dimension). The same approach applies to the occurrence coefficients $\{\bfalpha_k, \bfgamma_k, \bfomega_k\}$. Settings in which the number of samples is far smaller than the number of taxa typify high-dimensional sparse applications \citep{hmp2012}, and they are exactly where ARC is most advantageous.

\section{Simulation Study}
\label{sec:simulation-study}
 
Our simulation study asks whether the model can jointly recover two things: the active-set process that generates exact zeros and the latent positive dependence among components. That joint recovery is what separates ARC from models that treat zeros and dependence in isolation. Four goals motivate the study: (i) recovery of active-set structure at the observation-component level (classification of each cell $(i,k)$ as active or structurally zero), the component level (each component's prevalence $\frac{1}{n}\sum_i Z_{ik}$ aggregated across observations), the sample level (each observation's active count $m_i = \sum_k Z_{ik}$ aggregated across components), and the pairwise level (each pair's co-activation rate $\frac{1}{n}\sum_i Z_{ik} Z_{i\ell}$ aggregated across observations); (ii) recovery of abundance-scale latent dependence, set against recovery of the empirical observed-compositional correlation; (iii) performance across dimension and sparsity; and (iv) sensitivity to the dependence regime.
 
The study comprises ten replicates for each combination of sparsity regime, compositional dimension, and dependence structure described below. Each replicate has $n = 350$ observations and $d \in \{20, 50, 100, 250, 500, 1000\}$ compositional components. Every observation carries covariates $\bfw_i$ and a basis vector $\bfh_i$ built from radial basis functions centered at random subsets of simulated locations and standardized column-wise. We generated the active set from a probit latent-factor model by simulating $e_{ik}\sim \text{N}(0,1)$, setting
\[
    Z_{ik}^{\ast}
    = \bfw_i'\bfalpha_k + \bfh_i'\bfgamma_k + \bfomega_k'\bm{o}_i + e_{ik},
\]
and defining $Z_{ik} = \mathbf{1}(Z_{ik}^{\ast} > 0)$ with active probability
\[
    \rho_{ik}
    = \Phi\!\left(\bfw_i'\bfalpha_k + \bfh_i'\bfgamma_k + \bfomega_k'\bm{o}_i\right),
\]
where $\Phi$ is the standard normal distribution function. The occurrence factors $\{\bm{o}_i\}$ induced dependence among the active indicators such that components with similar occurrence loadings tended to appear together. Sparsity was controlled through the active-set intercept, sampled as $\alpha_k \sim \text{N}(\mu_\alpha, 0.35^2)$ for $k = 1, \ldots, d$ with $\mu_\alpha = 0.70$, $-0.15$, and $-0.95$ giving ``low," ``moderate," and ``high" sparsity. We also enforced a minimum of two active components per observation to avoid degenerate directional observations. When fewer than two were drawn active, the two components with the largest active probabilities were activated; the same rule applied to the posterior predictive active sets, which kept the simulation and the predictive evaluation consistent.
 
Conditional on the active set, abundance-scale latent active vectors $\bfx_i^+$ were drawn from the factor model. For active component $k\in A_i$, we simulated $\epsilon_{ik}\sim \text{N}(0,\sigma_k^2)$ and set
\[
    x_{ik}
    = \bfw_i'\bfbeta_k + \bfh_i'\bfdelta_k + \bff_i'\bflambda_k + \epsilon_{ik},
\]
with $\bff_i\sim \text{N}_q(\bfzero,\bfI_q)$ and $\bflambda_k$ fixing the latent dependence structure. Normalization gave the positive subcomposition
\[
    y_{ik}
    = \frac{x_{ik}}{\left(\sum_{\ell\in A_i}x_{i\ell}^2\right)^{1/2}},
    \qquad k\in A_i,
\]
and $y_{ik} = 0$ for $k\notin A_i$ such that the observed composition was $u_{ik} = y_{ik}^2$. The fitted model saw only the compositions $\{\bfu_i\}$ and the active set implied by exact zeros; the latent radii and the unnormalized vectors $\{\bfx_i^+\}$ were hidden.
 
The true latent covariance and correlation were $\bfV = \bfLambda\bfLambda' + \bfD$ and $\bfR = \diag(\bfV)^{-1/2}\bfV\diag(\bfV)^{-1/2}$, respectively, where $\bfLambda \in \R^{d \times q}$ is the factor-loading matrix of rank $q = 3$ and $\bfD = \diag(\sigma_1^2,\ldots,\sigma_d^2)$ collects the uniquenesses $\sigma_k^2 \sim \text{Unif}(0.035, 0.060)$ for $k = 1, \ldots, d$. We examined two regimes for $\bfLambda$. The block regime assigns components to $G = 4$ latent groups with within-group loadings $\bflambda_k = 0.70\,\bm{P}_{g(k)} + 0.08\,\bm{e}_k$, where $\bm{P}_g \in \R^q$ is a group-specific prototype and $\bm{e}_k \sim \text{N}_q(\bfzero, \bfI_q)$; the result is visually interpretable blocks of positive within-group correlation. The unstructured regime instead draws $\tilde{\bflambda}_k \sim \text{N}_q(\bfzero, 0.45^2 \bfI_q)$ and row-normalizes through $\bflambda_k = 0.45\sqrt{q}\,\tilde{\bflambda}_k / \|\tilde{\bflambda}_k\|$ such that each row has equal norm. We treat the unstructured regime as the realistic case, in which components share environmental drivers in graded, overlapping ways rather than partitioning into discrete groups. The block regime appears in Appendix B as a case whose latent structure can be visually inspected directly.
 
\subsection{Evaluation Metrics}
 
Recovery was assessed for both the active-set process and the latent positive-subcomposition dependence with active-set recovery summarized at four levels. Observation-component-level discrimination, the model's ability to classify each pair $(i, k)$ as active or structurally zero, was measured by the active-set area under the receiver operating characteristic curve (AUC), computed from posterior mean active probabilities $\widehat\rho_{ik}$ \citep{hanley1982meaning}. The AUC is threshold-free and stays interpretable when the prevalence of active entries varies across regimes; balanced accuracy near the threshold $\widehat\rho_{ik} > 0.5$ averages sensitivity and specificity and is preferable to true-zero recovery, which merely rewards matching the marginal zero rate. Component-level recovery was judged by the agreement between estimated and true component prevalence and sample-level recovery by the posterior predictive distribution of active counts. For the pairwise level, we estimated co-activation
\[
    C_{k\ell}^{Z}
    = \frac{1}{n}\sum_{i=1}^n
    \mathbf{1}(Z_{ik}=1,\ Z_{i\ell}=1)
\]
by $\widehat C_{k\ell}^{Z}= \frac{1}{n}\sum_{i=1}^n\widehat\rho_{ik}\widehat\rho_{i\ell}$ and report $\Corr\{\offdiag(\widehat\bfC^{Z}),\offdiag(\bfC^{Z})\}$, which captures whether the model recovers which component pairs tend to be active together rather than only the global zero rate.
 
Latent dependence recovery was summarized by $\Corr\{\offdiag(\widehat\bfR),\offdiag(\bfR)\}$, where $\widehat\bfR = \diag(\widehat\bfV)^{-1/2}\widehat\bfV\diag(\widehat\bfV)^{-1/2}$ and $\widehat\bfV$ is the posterior mean of $\bfV=\bfLambda\bfLambda' + \bfD$. We complemented this with the positive sign-recovery rate $P\!\left\{\widehat R_{k\ell} > 0.05\mid R_{k\ell} > 0.05\right\}$, which targets the claim of recovering positive latent dependence directly; the $0.05$ threshold keeps numerically negligible correlations from counting as substantive. We also report an observed-compositional-correlation baseline $\Corr\{\offdiag(\Corr(\bfU)),\offdiag(\bfR)\}$, which measures how much information about latent dependence the empirical correlation matrix of the observed compositions actually carries. This baseline is not a competing model; it summarizes how closure obscures information and is the natural benchmark against which the model's latent recovery should be judged.
 
\subsection{Joint Recovery under Realistic Latent Dependence}
 
Table \ref{tab:core-unstr} reports active-set recovery across dimensions and sparsity regimes. Discrimination between active sets was consistently strong: AUC between $0.83$ and $0.86$, balanced accuracy between $0.67$ and $0.76$, and pairwise co-activation correlation at least $0.97$ in every cell, confirming that we recovered the structure of co-occurrence rather than merely the marginal zero rate. The same table shows that the positive subcomposition layer recovered the latent correlation structure on every metric. Latent correlation was at least $0.93$ across all cells and above $0.99$ in many, and positive sign recovery was at least $0.89$. In contrast, the observed-compositional baseline fell sharply as sparsity increased, reaching $0.21$ under high sparsity at $d = 20$. That decline reflects the inherent behavior of raw compositional correlations under simplex closure, not a deficiency of the model. The gap between latent recovery and the observed baseline widens precisely where the method is most useful.
 
\begin{table}[ht]
\centering
\caption{Simulation results under unstructured latent dependence. AUC and balanced accuracy (Bal. acc.) summarize observation-component-level active-set discrimination using posterior mean active probabilities. Pair co-act.\ is the correlation between estimated and true off-diagonal pairwise co-activation rates. Latent corr.\ is the correlation between estimated and true off-diagonal latent correlations. Pos.\ sign is the recovery rate for truly positive latent correlations beyond threshold $0.05$. Obs.\ baseline is the correlation between the empirical observed-compositional correlation and the true latent correlation. The gap between Latent corr.\ and Obs.\ baseline summarizes how much closure obscures.}
\label{tab:core-unstr}
\begin{tabular}{llrrrrrr}
\toprule
Sparsity & $d$ & AUC & Bal.\ acc. & Pair co-act. & Latent corr. & Pos.\ sign & Obs.\ baseline \\
\midrule
low & 20 & 0.858 & 0.717 & 0.996 & 0.991 & 0.957 & 0.689 \\
low & 50 & 0.840 & 0.703 & 0.996 & 0.992 & 0.963 & 0.655 \\
low & 100 & 0.850 & 0.724 & 0.996 & 0.991 & 0.968 & 0.626 \\
low & 250 & 0.847 & 0.720 & 0.997 & 0.990 & 0.958 & 0.643 \\
low & 500 & 0.838 & 0.711 & 0.996 & 0.993 & 0.965 & 0.652 \\
low & 1000 & 0.843 & 0.716 & 0.996 & 0.997 & 0.976 & 0.659 \\
moderate & 20 & 0.848 & 0.758 & 0.990 & 0.990 & 0.953 & 0.433 \\
moderate & 50 & 0.826 & 0.738 & 0.991 & 0.984 & 0.953 & 0.471 \\
moderate & 100 & 0.836 & 0.748 & 0.993 & 0.990 & 0.957 & 0.482 \\
moderate & 250 & 0.845 & 0.757 & 0.993 & 0.992 & 0.963 & 0.510 \\
moderate & 500 & 0.835 & 0.747 & 0.993 & 0.993 & 0.966 & 0.527 \\
moderate & 1000 & 0.830 & 0.743 & 0.993 & 0.996 & 0.975 & 0.543 \\
high & 20 & 0.845 & 0.668 & 0.978 & 0.930 & 0.891 & 0.207 \\
high & 50 & 0.860 & 0.710 & 0.987 & 0.977 & 0.942 & 0.262 \\
high & 100 & 0.848 & 0.688 & 0.983 & 0.976 & 0.944 & 0.284 \\
high & 250 & 0.853 & 0.700 & 0.986 & 0.986 & 0.954 & 0.325 \\
high & 500 & 0.853 & 0.700 & 0.986 & 0.982 & 0.954 & 0.303 \\
high & 1000 & 0.852 & 0.699 & 0.988 & 0.991 & 0.965 & 0.350 \\
\bottomrule
\end{tabular}
\end{table}
 
\begin{figure}[h]
\centering
\includegraphics[width=\linewidth]{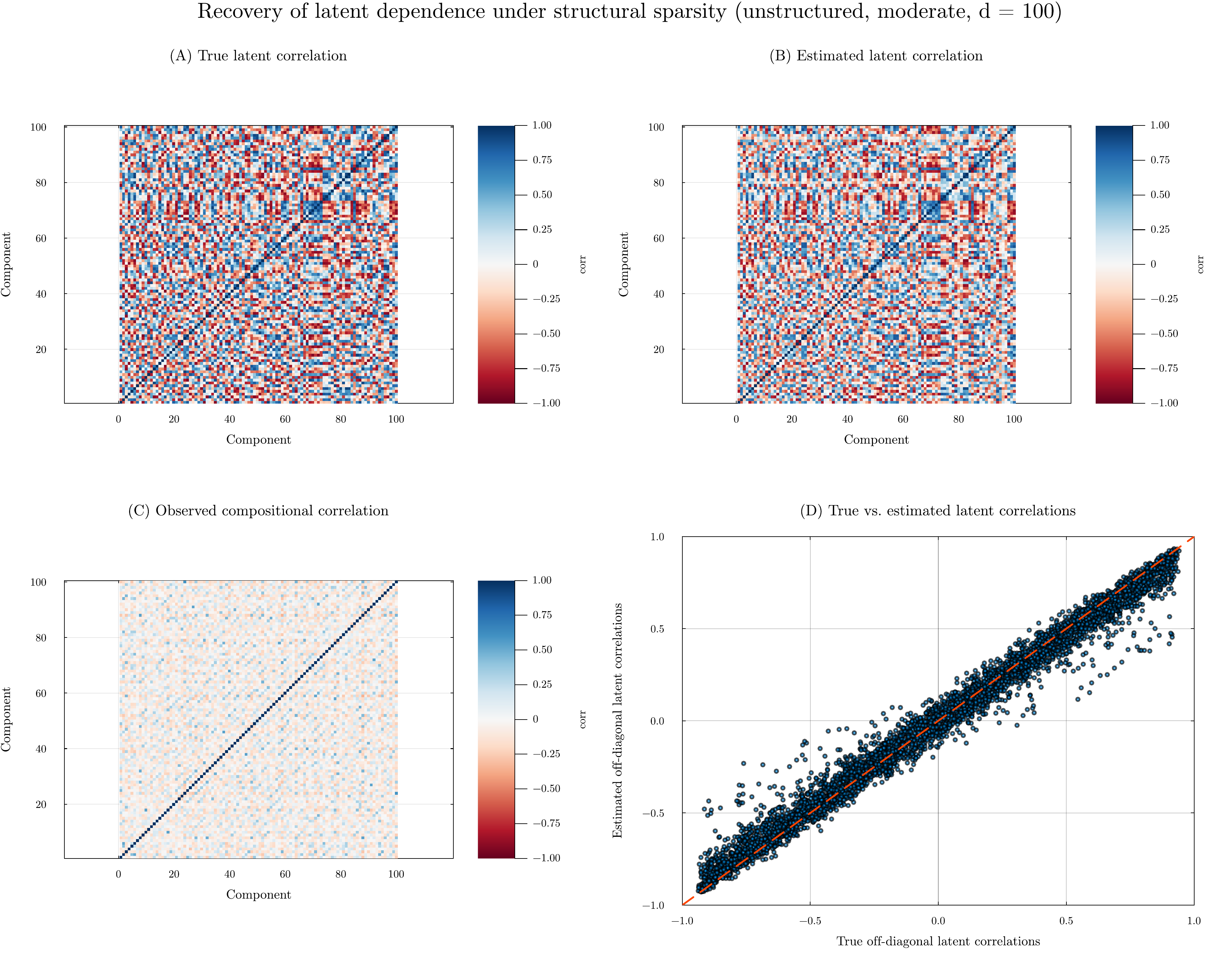}
\caption{Dependence recovery under unstructured factor covariance dependence with moderate sparsity at $d=100$. (A) True latent correlation $\bfR$. (B) Posterior mean latent correlation $\widehat\bfR$. (C) Empirical observed-compositional correlation $\Corr(\bfU)$, which carries little signal about the latent dependence structure. (D) Estimated against true off-diagonal latent correlations; the dashed line is the identity.}
\label{fig:joint-unstr-d100}
\end{figure}

Figure \ref{fig:joint-unstr-d100} depicts the unstructured, moderate sparsity, $d=100$ case. The posterior mean estimate (panel B) coincides with the true latent correlation (panel A), whereas panel C, the empirical observed-compositional correlation, is nearly indistinguishable from noise after closure. This contrast is central to the paper: positive latent co-response can persist even when raw correlations among observed proportions carry almost no interpretable signal about the latent dependence. Active-set recovery graphics for the same setting appear in Appendix C.
 
Active-set recovery stayed strong across sparsity regimes, though the interpretation of threshold-based metrics shifts with the zero rate, which is why we report AUC and balanced accuracy rather than raw true-zero recovery. Latent dependence recovery is most difficult when active sets are small because each pairwise latent association is informed only by observations in which both components are active. The factor structure partly relieves this bottleneck by borrowing strength across components.
 
Two patterns in Table \ref{tab:core-unstr} deserve comment. First, latent correlation and positive sign recovery are weakest at the smallest dimension and strengthen as $d$ grows with the weakest cell at high sparsity and $d=20$. Higher sparsity reduces the number of observations informing each pairwise association; the $\binom{d}{2}$ pairwise correlations are generated by only $q$ shared factors and become increasingly over-determined by a common factor subspace as $d$ grows. Each additional component carries more information about the same $q$ latent directions, and this borrowing of strength offsets the per-pair bottleneck that sparsity creates. Thus, the factor subspace and $\bfR$ are estimated more precisely at large $d$ even while individual active sets remain small; this is similar to the blessing-of-dimensionality phenomenon that \citet{chattopadhyay2024blessing} established for covariance models of the form $\bfV = \bfLambda\bfLambda' + \bfD$.
Holding the factor rank fixed, larger dimension makes more components load on the same factors and sharpens estimation of the shared subspace and the covariance; conditional on the radii and active sets, the ARC positive-subcomposition layer is exactly such a Gaussian factor model, so the same effect exists here with the active-set process restricting each observation to its active submatrix. The second pattern is that the active-set discrimination metrics are stable across $d$, which parallels the improving posterior concentration of inclusion-type rules in sparse high-dimensional multiple testing \citep{datta2013asymptotic}.
 
The simulation study supports three conclusions. First, the active-set model recovers exact-zero structure at several resolutions from observation-component-level active probabilities (the per-cell $(i,k)$ probability that component $k$ is active in observation $i$) to pairwise co-activation (the rate at which components $k$ and $\ell$ are simultaneously active across observations). Second, ARC recovers latent positive and negative dependence even when the observed compositional correlations are heavily attenuated by simplex closure. Third, the active set and the latent dependence are recovered jointly on the same sparse high-dimensional data, which is what sets the framework apart from models that handle zeros and dependence separately. Throughout, the target is the latent correlation matrix $\bfR$ implied by ARC before closure, not the empirical correlation matrix $\Corr(\bfU)$ of observed proportions, in which positive latent dependence is obscured.
 
 
\section{Discussion}
\label{sec:discussion}
 
We developed the Active-set Ray-based Compositional (ARC) framework for sparse, high-dimensional compositional data with structural zeros and latent dependence. A composition is mapped to the nonnegative orthant of the unit hypersphere, structural zeros are handled by an active-set process, and the positive subcomposition is modeled by evaluating a latent Gaussian density along the positive rays of the active subspace. This construction retains the latent Gaussian interpretation of projected Gaussian models, while avoiding the orthant normalizing constant that constrains truncated directional approaches. We implement the framework with a factor-augmented sampler, whose conditionally Gaussian updates facilitate high-dimensional inference and yield a tractable joint model for zeros and arbitrary dependence.
 
The proposed framework brings together four properties that are difficult to accommodate at once: exact zeros, flexible latent dependence, high-dimensional scalability, and compositional geometry. The active-set process determines which components are present, the ARC density governs the relative direction of the positive subcomposition, and the low-rank covariance captures arbitrary dependence on the abundance scale. In contrast, a model that substitutes small values for zeros or that uses a single continuous density to account for both presence and dominance blurs mechanisms that are genuinely distinct. Therefore, the contribution here is conceptual as much as it is computational: a composition is regarded as the observed direction of a latent abundance vector whose magnitude is unobserved and whose support is fixed by an explicitly modeled active set. This reframing is what allows structural zeros, flexible dependence, and high-dimensional scalability to coexist in one model, and it is what makes the approach well-suited for microbial, ecological, and environmental settings in which zeros and positive co-responses are scientifically central rather than nuisances.
 
The ARC model also clarifies the role of positive dependence in compositional data. On the observed simplex, the unit-sum constraint induces competition such that raw observed correlations may appear weak or even oppositely signed when the components in fact respond positively to shared latent drivers. ARC instead recovers the abundance-scale dependence that exists before closure, where two components may be positively correlated through a common response to an environmental gradient even while their observed proportions compete for a fixed total. The simulations make the gap concrete. Across every setting, the latent abundance correlation was at least $0.93$, whereas the observed-compositional baseline was far weaker and degraded sharply with sparsity.
 
The proposed construction also overcomes the principal computational obstacle to clipped projected Gaussian models. Truncation to the positive orthant is conceptually appealing but introduces a parameter-dependent orthant probability whose evaluation in high dimensions is computationally prohibitive. 
By enforcing support directly along the observed ray, the ARC model eliminates any rejection step or orthant normalizing constant. The composition-as-direction framing is beneficial for compositional data analysis for several reasons: (1) the observed quantities are relative directions and exact zeros, not rejected Gaussian vectors; (2) the latent radius represents the unobserved magnitude; and (3) the active set represents structural sparsity. The closest analogues to ARC either retain closure-based transformations and require positivity or retain truncated projected Gaussian likelihoods and require expensive orthant evaluations.
 
We adopt the low-rank factor form in \eqref{eq:lowrank} for its scalability and its interpretation in terms of shared latent gradients; however, this factor covariance is just one route to flexible dependence. An equally compatible alternative places sparsity directly on the precision matrix $\bfV^{-1}$, which yields a Gaussian graphical model whose nonzero entries encode conditional dependence among components on the abundance scale \citep{ha2020compositional, xu2026bayesian}. Because the ARC likelihood is conditionally Gaussian in $\bfx_i^+$ given the radii and active sets, such a precision-based specification can be substituted for the factor covariance without altering the construction. 

Two features of the factor representation deserve attention. Under weakly informative Gaussian priors on $\{\bflambda_k\}$ and a fixed factor rank $q$, the posterior tends to attenuate the strongest latent correlations toward zero at finite sample sizes, which is a well-documented property of factor analysis models \citep{anderson1956statistical, bhattacharya2011sparse}. The posterior mean of $\bfR$ still recovers correlation structure with high fidelity, though credible intervals at the highest-magnitude pairs may be conservative. A second consideration is our use of a fixed $q$. When an appropriate rank is not known \textit{a priori}, a prior that increasingly penalizes higher-order columns of $\bfLambda$, such as the multiplicative gamma process of \citet{bhattacharya2011sparse}, allows the effective number of factors to be learned adaptively. Such a cumulative shrinkage prior should widen the credible intervals at the highest-magnitude pairs and is compatible with the factor-augmented sampler without modification.
 
Several extensions are natural in the ARC framework. In particular, a spatial or spatio-temporal extension would be relevant in many scientific disciplines, where the latent mean
\[
    \bfmu_i = \bfB\bfw_i + \bfeta(\bfs_i,t_i)
\]
contains a multivariate spatial or spatio-temporal random effect $\bfeta(\bfs_i,t_i)$.
Such an extension is methodologically natural under the ARC framework but warrants its own treatment because it raises additional issues of spatial prediction, spatial confounding, and scalable Gaussian process computation. In high-dimensional applications, structure on $\bfeta$ is typically imposed through separable covariance \citep{genton2007separable}, basis expansions \citep{cressie2008fixed, wikle2010low}, dynamic factor models \citep{lopes2011generalized}, or multivariate Gaussian Markov random fields \citep{rue2005gaussian}. The ARC likelihood is compatible with such choices because, conditional on radii and active sets, the model becomes Gaussian in the augmented active-set vectors $\bfx_i^+ = r_i\,\bfy_i^+$. Another extension may include a detection model for sampling zeros in which a latent occupancy process is separated from the observed detection process. Then, the ARC model can be applied to the latent positive components, and the active-set layer can accommodate false negatives. 

The appendices contain the full Bayesian hierarchical model (Appendix A), an additional simulation study under a block-structured dependence regime (Appendix B), and active-set recovery diagnostics (Appendix C).
 
\section*{Acknowledgments}
 
We thank the Virginia Tech CoDA reading group members for their thought-provoking discussion: Rebecca Catlett, Jacob Gareis, Charlie Keglovitz, Andrew Pope, Dylan Steberg. We would also like to thank Brian D. Badgley, with whom we intend to collaborate for a future iteration of this manuscript containing a microbiome case study.
 
\bibliographystyle{plainnat}
\bibliography{references}
 
\section*{Appendix A - Bayesian Hierarchical Model}\label{app:model}
 
We fit the ARC model using the following Bayesian hierarchical framework:
\begin{align*}
    Z_{ik} &= \mathbf{1}(Z_{ik}^* > 0),\\
    Z_{ik}^* &\sim \text{N}\!\left(\bfw_i'\bfalpha_k + \bfh_i'\bfgamma_k + \bfomega_k'\bm{o}_i,\,1\right),\\
    \bfx_i^+=r_i\bfy_i^+ &\sim \text{N}_{m_i}\!\left(\bfB\bfw_i + \bfDelta\bfh_i + \bfLambda\bff_i,\,\bfD\right),  \quad \text{restricted to } A_i\\
    \bfbeta_k &\sim \text{N}_p(\bfbeta_0,\,25\,\bfI_p),\quad \bfbeta_0 = (1,0,\ldots,0)',\\
    \bfdelta_k &\sim \text{N}_L(\bfzero,4\,\bfI_L),\\
    \bflambda_k &\sim \text{N}_q(\bfzero,4\,\bfI_q),\\
    \bff_i &\sim \text{N}_q(\bfzero,\bfI_q),\\
    \sigma_k^2 &\sim \text{IG}(2,\,0.10),\\
    \bfalpha_k &\sim \text{N}_p(\bfzero,9\,\bfI_p),\\
    \bfgamma_k &\sim \text{N}_L(\bfzero,4\,\bfI_L),\\
    \bfomega_k &\sim \text{N}_{q_o}(\bfzero,4\,\bfI_{q_o}),\\
    \bm{o}_i &\sim \text{N}_{q_o}(\bfzero,\bfI_{q_o}),
\end{align*}
where $Z_{ik}^*$ is the Albert-Chib latent variable for the probit active-set model, $\bfalpha_k\in\R^p$ and $\bfgamma_k\in\R^L$ are the occurrence covariate and basis coefficients, $\bm{o}_i\in\R^{q_o}$ is the occurrence factor for observation $i$, $\bfomega_k\in\R^{q_o}$ is the occurrence loading for component $k$, $\bfy_i^+$ is the active subvector of the square-root composition, $r_i > 0$ is the latent radius, $\bfB = [\bfbeta_1,\ldots,\bfbeta_d]'\in\R^{d\times p}$ collects component-specific covariate coefficients, $\bfDelta = [\bfdelta_1,\ldots,\bfdelta_d]'\in\R^{d\times L}$ collects basis coefficients, $\bfLambda = [\bflambda_1,\ldots,\bflambda_d]'\in\R^{d\times q}$ collects positive subcomposition factor loadings, $\bff_i\in\R^q$ is the positive subcomposition factor for observation $i$, $\bfD = \diag(\sigma_1^2,\ldots,\sigma_d^2)$ collects the uniquenesses, $q$ is the rank of the positive subcomposition factor model, $q_o$ is the rank of the occurrence factor model, and $r_i$ is updated via random-walk Metropolis on $\log r_i$ as described in Section \ref{sec:implementation}. The prior mean $\bfbeta_0 = (1,0,\ldots,0)'$ assigns the intercept of each $\bfbeta_k$ a prior mean of one rather than zero, which weakly anchors the latent abundance scale during sampling. We resolve the multiplicative scale invariance of the likelihood noted in Section \ref{sec:ray-model} by the \textit{post hoc} rescaling of Section \ref{sec:implementation}, which affects neither the identifiable correlation matrix $\bfR$ nor the active probabilities $\bm{\rho}$.
 
\newpage 
 
\section*{Appendix B - Block-Structured Dependence as a Visualization}
 
\begin{figure}[ht]
\centering
\includegraphics[width=0.95\linewidth]{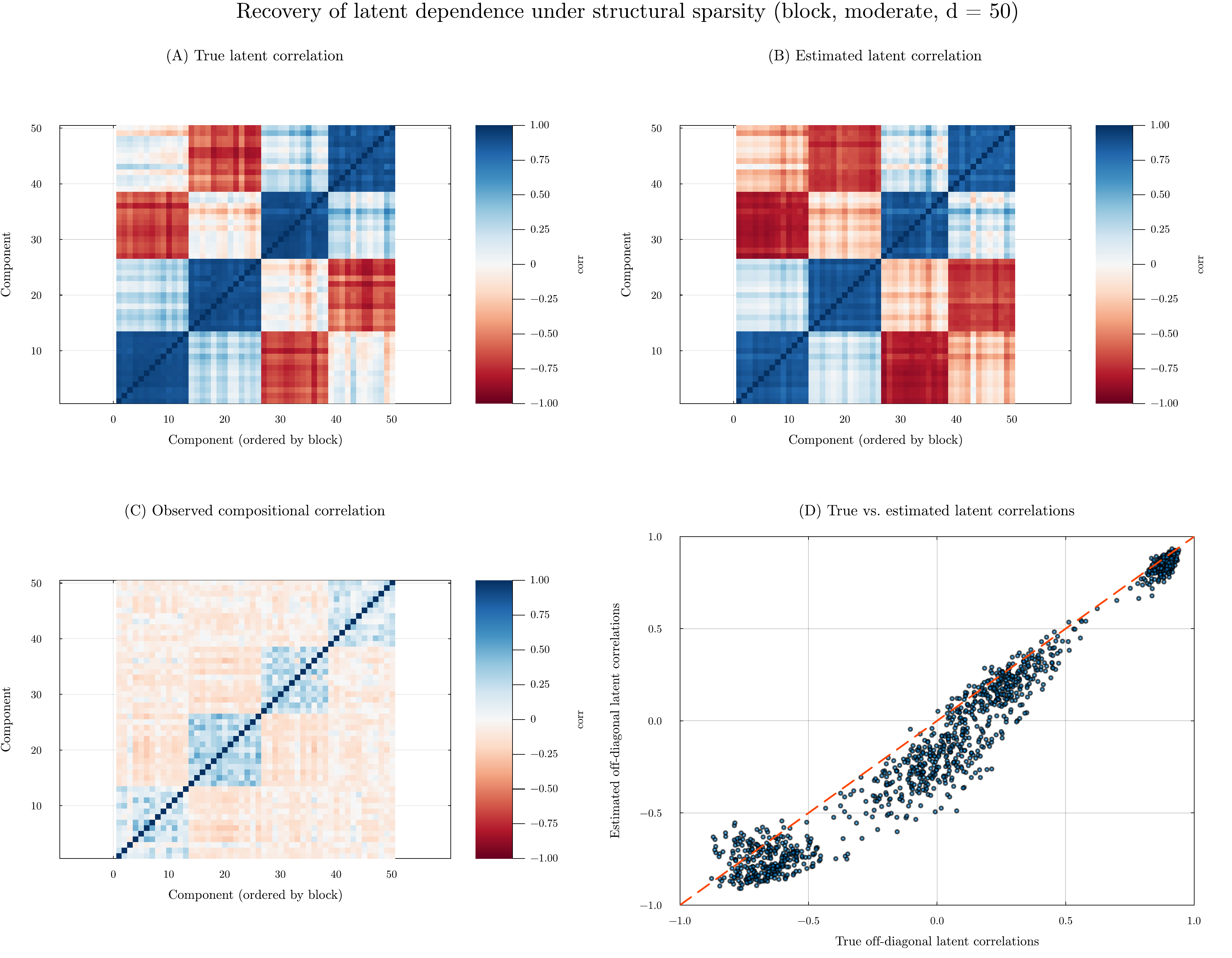}
\caption{Dependence recovery under block-structured dependence with moderate sparsity at $d=50$. Panels are as in Figure \ref{fig:joint-unstr-d100}. The block organization permits the latent dependence structure to be inspected visually, whereas the closure-obscured empirical correlation in panel~(C) is substantially attenuated.}
\label{fig:joint-block-d50}
\end{figure}

To complement the unstructured analysis, we repeat the simulation under a block-structured covariance regime in which components are partitioned into latent groups with similar within-group loadings. The block, moderate-sparsity, $d=50$ case is depicted in Figure \ref{fig:joint-block-d50}. Panel (B) shows the posterior mean estimate, which closely reproduces the true block pattern in panel (A). Panel (C) shows the empirical observed-compositional correlation, which is substantially attenuated after closure, and panel (D) shows that the estimated off-diagonal correlations concentrate around the identity. The block regime is a stylized setting, but the visual contrast between panel (B) and panel (C) makes the closure-induced obscuring of latent dependence clear.
 
Quantitative results for the block regime at $d \in \{20, 50, 100\}$ are reported in Table \ref{tab:core-block}. The pattern coincides with the unstructured results at comparable dimensions: AUC between $0.84$ and $0.88$, balanced accuracy between $0.71$ and $0.75$, pairwise co-activation correlation at least $0.98$, latent corr.\ at least $0.96$, and observed-compositional baseline degrading from approximately $0.80$ under low sparsity to $0.41$ at high sparsity and $d=20$. The block table is presented at moderate dimensions for visual interpretability rather than as a high-dimensional scaling argument.
 
\begin{table}[ht]
\centering
\caption{Simulation results under block-structured dependence at moderate dimensions. Columns are as in Table \ref{tab:core-unstr}. The block regime is presented for visual interpretability of the latent dependence structure; scaling to higher $d$ is reported in Table \ref{tab:core-unstr} under the unstructured regime.}
\label{tab:core-block}
\begin{tabular}{llrrrrrr}
\toprule
Sparsity & $d$ & AUC & Bal.\ acc. & Pair co-act. & Latent corr. & Pos.\ sign & Obs.\ baseline \\
\midrule
low & 20 & 0.869 & 0.747 & 0.995 & 0.991 & 0.889 & 0.820 \\
low & 50 & 0.858 & 0.728 & 0.997 & 0.989 & 0.855 & 0.802 \\
low & 100 & 0.843 & 0.713 & 0.996 & 0.990 & 0.837 & 0.772 \\
moderate & 20 & 0.843 & 0.753 & 0.991 & 0.985 & 0.812 & 0.658 \\
moderate & 50 & 0.843 & 0.754 & 0.993 & 0.988 & 0.862 & 0.717 \\
moderate & 100 & 0.836 & 0.748 & 0.992 & 0.985 & 0.803 & 0.716 \\
high & 20 & 0.877 & 0.726 & 0.985 & 0.971 & 0.795 & 0.413 \\
high & 50 & 0.861 & 0.709 & 0.985 & 0.984 & 0.832 & 0.540 \\
high & 100 & 0.864 & 0.715 & 0.988 & 0.969 & 0.780 & 0.568 \\
\bottomrule
\end{tabular}
\end{table}
 
\section*{Appendix C - Active-Set Recovery}
 
\begin{figure}[ht]
\centering
\includegraphics[width=\linewidth]{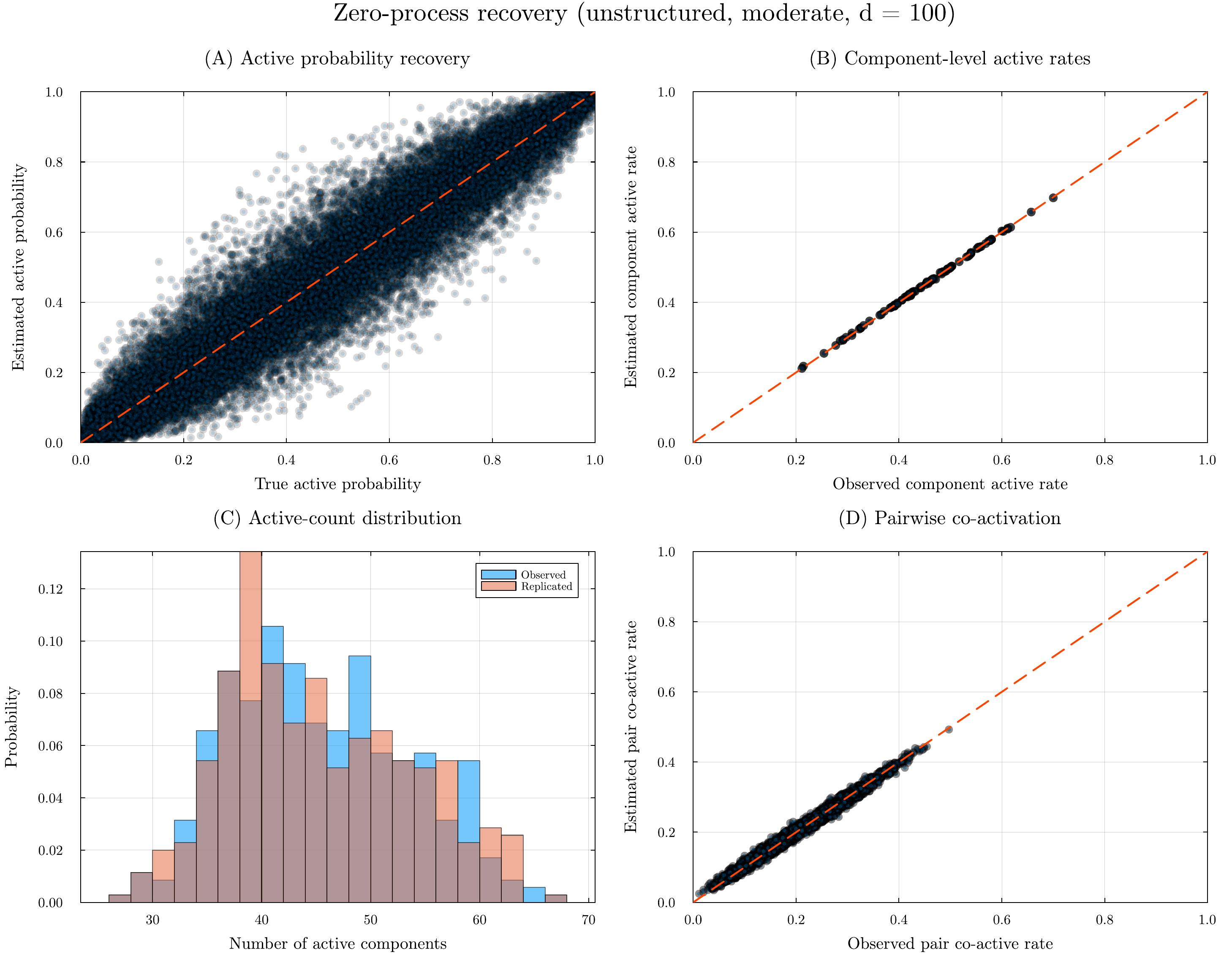}
\caption{Active-set recovery diagnostics under unstructured factor covariance dependence with moderate sparsity at $d=100$. (A) Posterior mean active probabilities $\widehat\rho_{ik}$ plotted against true active probabilities $\rho_{ik}$. (B) Posterior mean component prevalence $\frac{1}{n}\sum_i \widehat\rho_{ik}$ plotted against true component prevalence $\frac{1}{n}\sum_i Z_{ik}$. (C) Posterior predictive distribution of the per-observation active count $m_i = \sum_k Z_{ik}$ (replicated) overlaid on the empirical distribution (observed). (D) Posterior mean pairwise co-activation rate $\widehat{C}_{k\ell}^{Z}$ plotted against true pairwise co-activation rate $C_{k\ell}^{Z}$.}
\label{fig:zero-process-unstr-d100}
\end{figure}

\begin{figure}[ht]
\centering
\includegraphics[width=\linewidth]{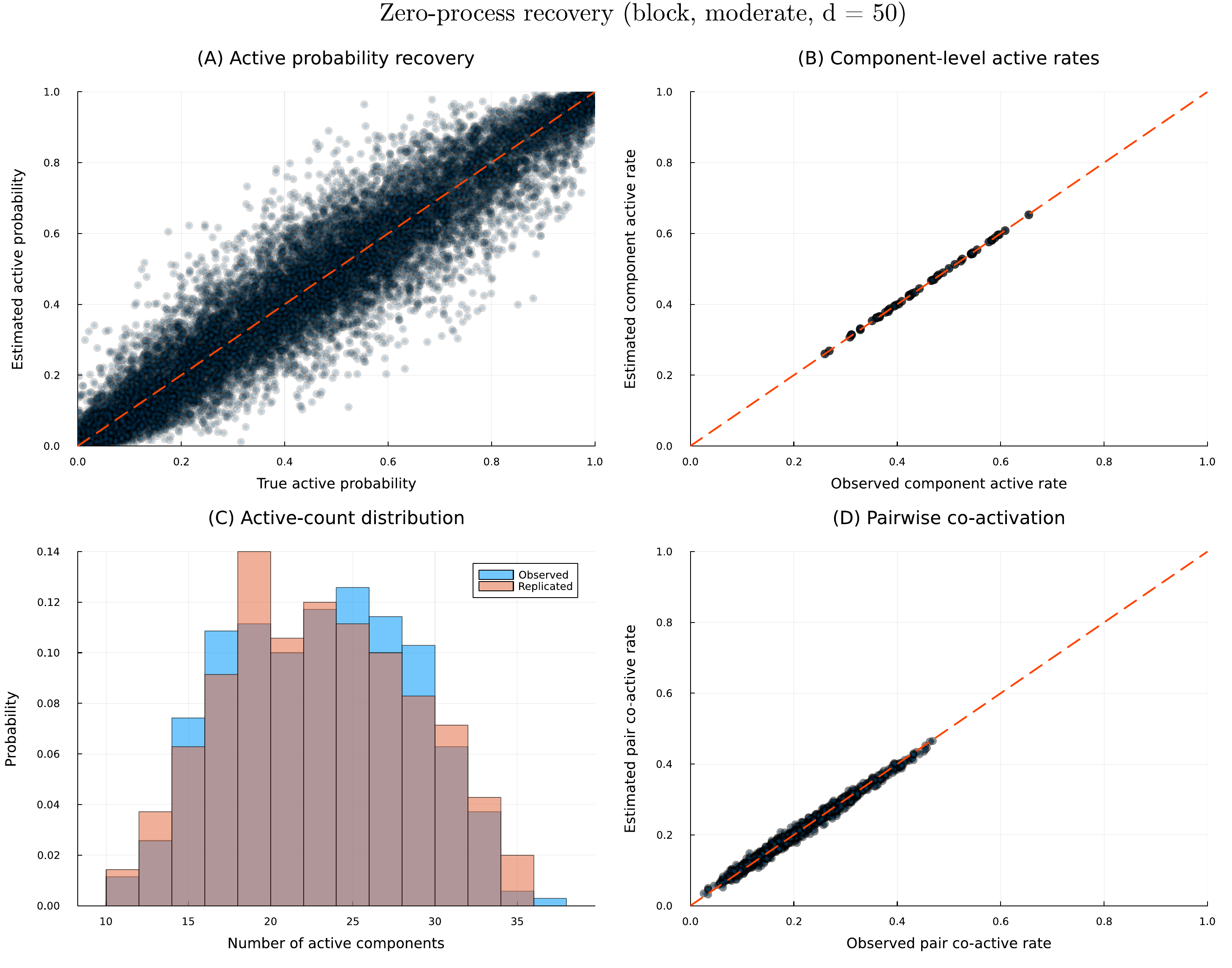}
\caption{Active-set recovery diagnostics under block factor covariance dependence with moderate sparsity at $d=50$. (A) Posterior mean active probabilities $\widehat\rho_{ik}$ plotted against true active probabilities $\rho_{ik}$. (B) Posterior mean component prevalence $\frac{1}{n}\sum_i \widehat\rho_{ik}$ plotted against true component prevalence $\frac{1}{n}\sum_i Z_{ik}$. (C) Posterior predictive distribution of the per-observation active count $m_i = \sum_k Z_{ik}$ (replicated) overlaid on the empirical distribution (observed). (D) Posterior mean pairwise co-activation rate $\widehat{C}_{k\ell}^{Z}$ plotted against true pairwise co-activation rate $C_{k\ell}^{Z}$.}
\label{fig:zero-process-block-d50}
\end{figure}

We present active-set recovery diagnostics that complement the latent correlation results in Section \ref{sec:simulation-study}. In particular, we depict observation-component-level active probability calibration, component-level prevalence agreement, sample-level active-count distributions, and pairwise co-activation recovery for the unstructured, moderate-sparsity, $d=100$ case (Figure \ref{fig:zero-process-unstr-d100}) and the block, moderate-sparsity, $d=50$ case (Figure \ref{fig:zero-process-block-d50}). The four panels in each figure depict the four resolutions of active-set recovery summarized in Table \ref{tab:core-unstr} and Table \ref{tab:core-block}, thereby providing visual confirmation of the quantitative metrics.
 
\end{document}